\documentclass{svmult}

\usepackage{amsmath}
\usepackage{url}
\usepackage{graphicx}

\begin{document}

\title*{On the Singlet Penguin in
$B\to K \eta'$ Decay\protect\footnotetext{Presented by K. Kumeri\v{c}ki at
9th Adriatic Meeting, Particle Physics and the Universe,
Dubrovnik, Croatia, September 4 - 14, 2003, to appear in the proceedings.}}
\author{Jan Olav Eeg\inst{1}\and
Kre\v{s}imir Kumeri\v{c}ki\inst{2} \and
Ivica Picek\inst{2}}
\institute{Department of Physics, University of Oslo, N-0316 Oslo, Norway
\and Department of Physics, Faculty of Science, University of Zagreb,
 P.O.B. 331, HR-10002 Zagreb, Croatia}
%
%
\maketitle

\begin{abstract}
New contributions to the short-distance $b\to s\eta'$ transition
are considered. They correspond to the quark and gluon content
of $\eta'$. Although substantially larger than the referent
QCD anomaly tail, they still cannot account for the observed
$\eta'$ enhancement.
\end{abstract}

Recent measurements by CLEO, BaBar and Belle collaborations
\cite{CLEO00,CLEO99,BABAR01,BELLE01,BELLE01b} 
of two-body charmless hadronic $B$ meson decays with $\eta'$ meson
in final state \cite{HFAG}
\begin{align}
Br(B^+\to K^+ \eta') &= (77.6 \pm 4.6) \cdot 10^{-6}\;, \nonumber \\
Br(B^0\to K^0 \eta') &= (65.2 \pm 6.0)\cdot 10^{-6} \;,
\label{ketap} 
\end{align}
indicate an enhancement by a factor of 5--6 when compared to the 
corresponding decays to the pion instead of  $\eta'$.
In an attempt to explain a dynamical origin of such an enhancement,
one can start from the observation that some exceptional properties
of $\eta'$ particle are related to the QCD anomaly. Namely,
the puzzle of the unexpectedly large $\eta'$ \emph{mass} (the famous U(1)
problem) was resolved by taking into account the QCD axial anomaly and the
pure gluonic component of the $\eta'$ quantum state
\begin{equation}
 | \eta' \rangle =\cdots +  | \mbox{gg} \rangle + \cdots \;.
\end{equation}
Thus, it was no surprise that in various dynamical mechanisms
invoked to explain the enhancement (\ref{ketap})
\cite{AtS97,HaZ97,KaP97,HoT97,DaHP97,DukY97,AhKS97,AlCGK97,KoS01,XiCL02,BeN02,FrZ03} this gluonic component often played a prominent role.

\begin{figure}
\center
\includegraphics[scale=0.7]{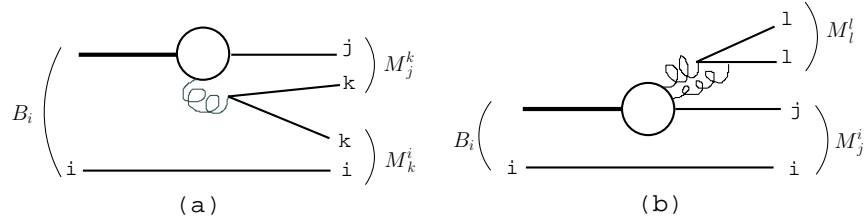}
\caption{\label{su3} Two flavour topologies contributing to
$B\to K \eta'$ decay in  SU(3) symmetry approach:
penguin (a) and  singlet-penguin (b). $B_i$ is $B$ meson triplet and
$M_{\hspace{1ex}i}^{j}$ is a pseudoscalar meson nonet. }
\end{figure}

On the other hand, more phenomenological analysis based on the SU(3)
flavour symmetry and the corresponding diagrammatical 
formalism \cite{ChGR03}, suggests
that besides the usual penguin (P) diagrams (Fig. \ref{su3}a) dominating
the $B\to K\pi$ rates, the
singlet penguin (SP) diagram (Fig. \ref{su3}b) should also contribute
substantially to the processes of the $B\to K \eta'$ type.
 Since the singlet penguin diagram can naturally be realized via 
creation of $\eta'$ particle by the pure gluonic intermediate state
(as already suggested on Fig. \ref{su3}b) this points to such 
gluonic mechanisms as
good candidates for explaining the enhancement for $\eta'$ modes (\ref{ketap}).

\begin{figure}
\center
\includegraphics[scale=1.0]{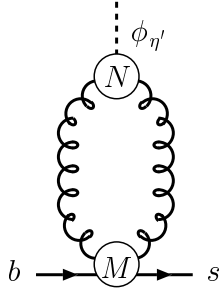}
\caption{\label{gluonloop}The hard gluon loop contribution to the
$b \to s \eta'$ transition.}
\end{figure}

In our recent paper \cite{EeKP03} we identified a well-defined short-distance
(SD) contribution to the singlet penguin amplitude, generating $b\to s\eta'$
transition displayed on Fig.~\ref{gluonloop}. This SD amplitude
corresponds to the hard gluons in the 
intermediate state producing a final $\eta'$ particle via a quark triangle
loop (represented by the blob $N$ in Fig.~\ref{gluonloop}), thus corresponding 
to a sort of a remnant of the QCD anomaly
(dubbed the ``anomaly tail'' in \cite{EeKP03}).

\begin{figure}
\center
\includegraphics*[scale=0.45]{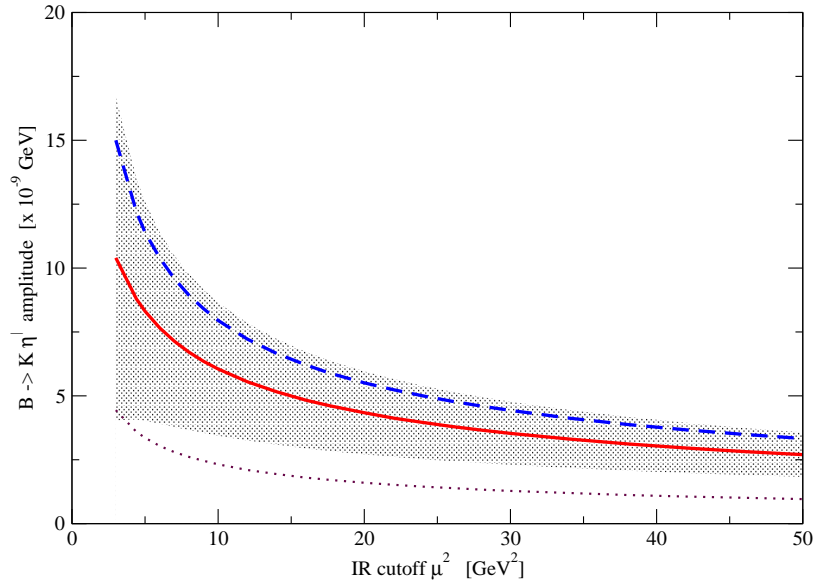}
\caption{\label{gg} Short-distance hard gluon contribution to
$B\to K\eta'$ amplitude. Dashed
line corresponds to pure quark content of $\eta'$ 
(\protect\ref{Fetagg}), while shaded area corresponds to allowed
region when also gluonic content of $\eta'$ is taken into account.
Dotted line is ``anomaly tail'' from \protect\cite{EeKP03}}
\end{figure}

Now we supplement this analysis by investigating further contributions
to the bi-gluonic $b\to s\eta'$ amplitude on Fig.~\ref{gluonloop} with
vertex $N$ represented by the expression
\begin{equation}
\label{fetagg}
 N_{\mu\nu}^{aa'}(\bar{Q}^2,\omega)=-i F_{\eta' g^* g^*}(\bar{Q}^2,\omega)
\epsilon_{\mu\nu k_1 k_2} \delta^{aa'} \;,
\end{equation}
where $\bar{Q}^2=-(k_{1}^2+k_{2}^2)/2$ is average virtuality of
gluons and $\omega=(k_{1}^2-k_{2}^2)/q^2$ is asymmetry parameter.
The form-factor $F_{\eta' g^* g^*}(\bar{Q}^2,\omega)$ for the symmetric
case $\omega=0$, and in the asymptotic limit valid
for large $\bar{Q}^2$ is given by \cite{AlP00,KrP02}
\begin{equation}\label{Fetagg}
  F_{\eta' g^* g^*}(\bar{Q}^2,0) =
   4\pi \alpha_s(\bar{Q}^2) \frac{f^{1}_{\eta'}}{\sqrt{3} \bar{Q}^2} \;,
\end{equation}
with $f^{1}_{\eta'}\approx 0.15$ GeV. Combining this with the 
amplitude $A_i(-Q^2)$ given in \cite{EeKP03} for the flavour-changing
 $b\to s g^* g^*$ transition (the blob $M$ in Fig.~\ref{gluonloop}),
we obtain the $b\to s\eta'$ amplitude
\begin{equation}
 \mathcal{A} (b\to s\eta')=\frac{G_F f_{\eta'}^{1}}{2\sqrt{6} \pi^2} 
\big(\phi_{\eta'}\bar{s}
 \gamma\cdot K L b\big) \sum_{i=u,c,t} \lambda_{i} \int
 dQ^2 \:\frac{\alpha_{s}^{2}(Q^2)}{Q^2}\: A_{i}(-Q^2) \;.
\label{one}
\end{equation}
Whereas the form-factor (\ref{Fetagg}) takes into account only transition to
the quark $|q\bar{q}\rangle$ quantum state of $\eta'$, it is interesting
to study the influence of gluonic $|gg\rangle$ component of $\eta'$ on
this result. It turns out \cite{KrP02} that
in the symmetric case of two gluons having similar momenta 
($\omega\approx 0$) the effect
of the $|gg\rangle$ component can be included by multiplying the
$\eta'g^{*}g^*$ form-factor (\ref{Fetagg}) by a factor
\[
    1-\frac{1}{12}B_{2}^{g} \;.
\]
Here, the range of allowed values for the Gegenbauer coefficient $B_{2}^{g}$
(obtained by a fit to the $\eta'$ transition form-factor)
can be found in \cite{KrP02}.
The error in $B_{2}^{g}$ is large, and for the most of the
allowed region the gluon contribution will interfere destructively, because
of the minus sign in the above factor. Accordingly, the amplitude will be
smaller than in the pure $|q\bar{q}\rangle$ case by an average of 30 percent.
This is displayed on Fig.~\ref{gg} where the quark transition
amplitude (\ref{one})
is combined with the spectator quark in order to
produce the physical $B\to K\eta'$ amplitude.
\vspace*{5ex}

Comparison to results of \cite{EeKP03} shows that the SD 
contributions considered
here are substantially larger than the ``anomaly tail'' part. Still,
they cannot explain the observed $\eta'$ 
enhancement (\ref{ketap}) by themselves.
Apart from some attempts to invoke new physics beyond the Standard
Model \cite{XiCL02},
another mechanism incorporating long distance aspects of the QCD anomaly
\cite{EeKP04} and/or the one of the penguin interference 
\cite{Li81,BeN02} seems to be needed to complete the picture.

\end{document}